\documentclass[twocolumn,showpacs,preprintnumbers,amsmath,amssymb]{revtex4} 
                                                                                
\usepackage{graphicx}
\usepackage{dcolumn}
\usepackage{bm}
\begin{document}
                                                                                
                                                                                
\title{Modelling temporal and  spatial features of collaboration network}
                                                                                
\author{Anjan Kumar Chandra}
\author{Kamalika Basu Hajra}
\author{Pratap Kumar Das}
\author{Parongama Sen}%
\affiliation{%
Department of Physics, University of Calcutta,92 Acharya Prafulla Chandra Road,
Calcutta 700009, India.\\
}%
                                                                                
\date{\today}

\begin{abstract}

The collaboration network is an example of a social network 
which has both non-trivial temporal and spatial dependence.  
Based on the observations  of collaborations in Physical Review Letters, 
a model of collaboration network is proposed which 
correctly reproduces the time evolution of the link length 
distributions, clustering coefficients, degree distributions 
and assortative property of real data to a large extent.
\end{abstract}
\pacs{87.23.Ge,89.75.Hc}
                                                                                
\def\be{\begin{equation}}
\def\ee{\end{equation}}
\maketitle



\section{Introduction}

Ever since the discovery of small world effect in a variety
of networks \cite{watts}, study of real world networks 
and their theoretical modelling have generated tremendous activity.
A network is equivalent to a graph and is characterised by the links which
connect pairs of nodes. 
Based on observations and theoretical arguments, it has been 
established that factors like preferential attachment, duplication, geographical distance, 
aging etc. are responsible in determining the connectivity in 
many real world networks \cite{BA}.

A scientific collaboration network is an example of a social network \cite{psreview} in
which the scientists are the nodes and a link is established between 
two scientists if they co-author a paper. 
Scientific collaboration networks of different types 
have been studied in detail for quite a few real databases 
\cite{newman1,newman2,ba_collab,inertia}. 
The average shortest distance in these networks
turns out to be of the order of $\log(N)$, where $N$ is the total 
number of authors  in these networks indicating that they form
small world networks. 
In fact, in \cite{ba_collab}, the data indicated that the average shortest 
distance might as well decrease with $N$ when $N$ is allowed to vary.
The degree distribution has a power law tail for many of the 
data bases, especially when the average degree is quite high, e.g., in the MEDLINE data \cite{newman1}. When the average degree is quite small,
e.g., in the  hep-th database, there is apparently an exponential cutoff.
There could also be two regimes of power law behaviour as argued 
in \cite{ba_collab}.
The clustering coefficient for collaboration networks is usually quite high
as any collaboration involving more than two authors will invariably contribute
to the clustering coefficient \cite{BA}. 
Collaboration networks are also known to have a positive assortativity \cite{psreview,assort}. 

Modelling of the collaboration network has been attempted in several
studies \cite{models} assuming a scheme where nodes are added regularly and they get attached to the existing nodes obeying a definite rule.
In these models, the effect of 
time and geographical distances are  usually ignored. 
Recent research has shown, on the other hand, that geographical distance
is a major factor determining scientific collaborations within or across countries 
\cite{katz,nagpaul,drift,olson}.
The question of distance dependence of links in real Euclidean networks 
has been addressed in several other real world networks \cite{yook,gastner,guimera,eguiluz,traffic}.
In collaboration networks, the link lengths are also expected to depend on time.
In earlier times, collaborations were largely limited between scientists 
located close to each other, as it required face to face  interactions 
to a considerable extent. The cost and inconvenience of communication
and travelling was largely responsible for collaborations taking  
place locally in majority of cases. With the communication undergoing 
rapid improvement  in the form of e-mails, electronic file transfer, 
fax, telephone etc.,
much larger number of  long distance collaborations 
have been possible  in more recent times. 

Thus it is expected that if the link length distributions in a scientific collaboration network
are calculated at different 
times, it would   show a noticeable change
with the  probability  of large link lengths increasing in time.
 A few theoretical models of networks
have  considered links to be distance dependent 
\cite{ps_rev}. 
A model for collaboration network should in principle
contain both distance and time dependent factors. 
However, one needs to have a
quantitative idea of the effect of time and distance in real collaboration networks. 
The  few studies which are  available \cite{katz,nagpaul,drift,olson}, do not give a 
complete picture. Since collaboration networks can be defined in a variety
of ways (e.g., based on a particular journal, based on one or more
scientific communities independent of any journal etc.) it is difficult
to speak of an absolute picture. In the present study we have 
attempted to model a particular kind of collaboration network
based on our observations of the data of Physical Review Letters (PRL) 
from the year 1965 to 2005.

In section II we discuss the 
real world data available so far and report in detail our own observations 
of the PRL data. In sec III, we discuss the model
and in section IV we summarise and discuss the results. 

\section {Real World Data} 

 In \cite{katz},  
 the distribution of geographical distances 
between co-authors  was studied restricting the studies to individual countries. The
results showed an exponential decrease with distance. The time evolution was
not studied in this case. Obviously  inter continent or inter country data 
were not considered here so the link lengths are restricted to a large extent.
The data was also up to 1990, when the communication revolution was yet to take 
shape.
In \cite{nagpaul}, it was shown that the geographical proximity 
has the greatest impact on transnational collaborations when
compared to other (thematic and socio-economic) factors.
In \cite{drift}, the link geographical distance between Economists sharing
publications was considered as an example to support a general model
of social network in the background of technological
advancement. Publications which had at   least one collaborator from the US
were considered only and data for the first two authors
of each paper were taken. The distance factor was also 
coarse grained. 
It was indeed found   that individual separations decrease with time. However,
the exact behaviour of the  distribution of link lengths was not presented possibly because of the  restricted nature of the data. 
In \cite{olson},  it was concluded that  
improvement of communication alone cannot help in 
long distance collaborations as there are other factors involved.

To obtain the link length distribution in a scientific collaboration network, 
one should,
in principle, take the network of collaborators of the particular type one
is interested in (e.g., physicists, economists or maybe even more specialised, like only
condensed matter physicists) and 
 calculate the 
 geographical distances separating them at the time of a collaboration.
However, it is more convenient to take a journal based data as has been done previously
in many studies.
We have therefore taken sample papers (at least 200 for each year for nine
different years between 1965 to 2005) randomly 
from the 
Physical Review Letters.  The task is then to  
 calculate the pairwise geographical distance between   
the host institutes of the authors  coauthoring a paper.
 However, since these addresses span the whole world,
this involves data over a wide range. Also,
the within city and within institute/university distances
are not readily available. We have therefore obtained the distance distribution in an indirect and coarse grained way which is described in the following. 

 To  author X in a paper we associate
the indices $x_1, x_2, x_3$ and $x_4$ ($x_i$'s are integers) which 
represent the 
University/Institute, city, country and continent of X respectively. 
Similar
indices $y_1, y_2, y_3$ and $y_4$ are defined for author Y. If, for
example, authors X and Y belong to the same institute, $x_i=y_i=1$ for
all $i$. On the other hand, if they are from different countries but
the from same continent, $x_4=y_4$ but $x_i \neq y_i$ for $i <  4$.  
We find out for what
maximum value of $k$,  $x_k \ne y_k$. The distance  between X and Y 
is then $l_{XY} = k +1$. If $x_i = y_i$ for all values of $i$ it means 
$l_{XY} =1$ according to our convention. 
As an example, one may consider the paper  PRL {\bf 64}  2870 (1990),
which features 4 authors. Here authors 1 and 2 are from the same 
institute in Calcutta, India, and are assigned  the variables 1, 1, 1, 1.
The 3rd  author belongs to a different institute in Calcutta and therefore
gets the indices 2, 1, 1, 1. The last author is from an institute in Bombay, India, 
and is assigned the variables 3, 2, 1, 1. 
Hence the pairwise distances are: $l_{12} = 1, l_{13}=l_{23}=2, l_{14}=l_{24}=l_{34}=3$.
The pair-wise distances $l$ gives the distribution $P(l)$ of
the distance between two collaborating authors.
We have also defined a distance factor $d$ for each paper where 
$d$ is the average of the pair-wise distances of authors coauthoring 
that paper. The corresponding  distribution $Q(d)$ has also been computed.
In the above example, 
the average  $d=2.333$.
Note that in $P(l)$, the fact that 
$l_{12}, l_{13}$ and $l_{23}$ are obtained from a single collaboration act 
is missing. Hence, in a sense, $Q(d)$ takes care of the correlation between the
distances. Let us call $Q(d)$ the correlated distance distribution.
Defining the distances in this way, the values of $l$ are 
discrete while the $d$ values have a continuous variation.
For papers with two authors, the two distributions are identical
but will be different in general.

In order to show that our coarse graining of distances is consistent
with the actual distances, we have picked up cities at random and plotted
the real distance $D$ against the coarse grained distance $l$ in Fig. 1.
Real data is available for $l=3,4$ and $5$ \cite{realdata} only as $l=1$ and $2$ correspond
to within institute and within city distances respectively. For these 
two $l$ values, we have assumed realistic average values of $D$, e.g.,  
$D \sim 0.1 $km for $l=1$ and $D\sim 10$ km for $l=2$ and shown in the 
same plot. We find that there is indeed a correlation between $D$ and $l$.
With only such a few points, it is difficult to ascertain the
exact behaviour of $D$ with $l$,  $D \sim \exp(\alpha l^\beta)$ 
with $\beta \sim 1$ could be a possible fit as shown in the figure.


\begin{center}
\begin{figure}
\noindent \includegraphics[clip,width= 5cm, angle=270]{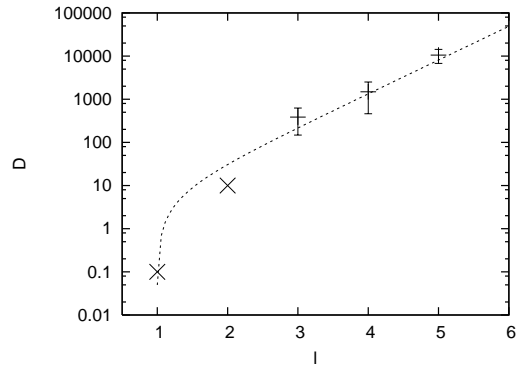}
\caption{ Relation between actual distance $D$ (in km) and coarse grained 
distance $l$. The data for $l=3,4$ and $5$ are from real distances
averaged over  50 to 100 data and for $l=1$ and $l=2$,
we assume some realistic average values of $0.1$ km and $10$ km (shown by $\times $ symbols)
respectively. The data fits fairly well to $1.5\exp(1.8l) -10$.  
               } 
\end{figure}
\end{center}

\begin{center}
\begin{figure}
\noindent \includegraphics[clip,width= 5cm, angle=270]{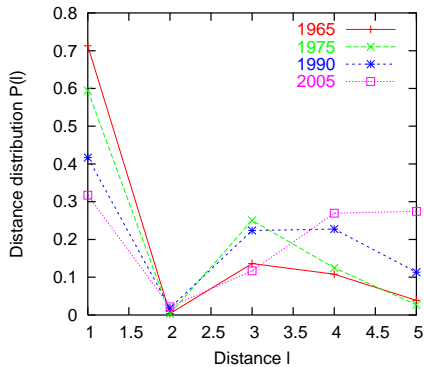}
\caption{Distance distribution $P(l)$ as function of distance $l$ for
              different years. } 
\end{figure}
\end{center}
\begin{center}
\begin{figure}
\noindent \includegraphics[clip,width= 5cm, angle=270]{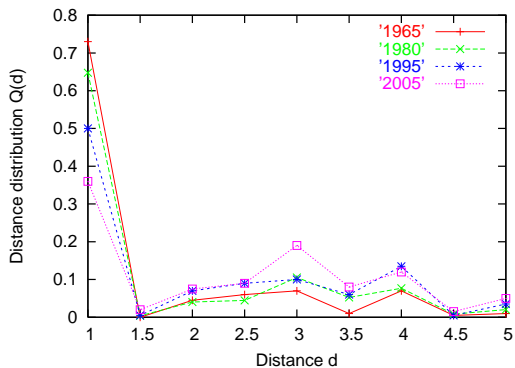}
\caption{Correlated distance distribution $Q(d)$ vs distance $d$ plot for 
different years are shown.}
\end{figure}
\end{center}

We have made exception for USA authors since it is a big country
comparable in size to Europe which consists of many countries. 
(Of course there are other big countries in the world but majority of
contributions to PRL are from the USA and Europe.)
 Thus  two authors belonging to, say, Kentucky and Maryland will have
 different country indices, i.e., 
 $x_3 \ne y_3$.

 Some papers like the experimental high energy 
 physics ones typically involve many authors and many institutes. We 
 have considered an upper bound, equal to 20, to
 the number of institutes and no bound for the number of authors.
 In case of multiple addresses, only the first one has been
 considered.

In Figs. 2 and 3, the distributions $P(l)$ and $Q(d)$ are shown. 
The two distributions have similar features  
but  differ in magnitude, more so in recent years, when the  
number of authors 
is significantly different from two in many papers.

Both the distributions  $P(l)$ and $Q(d)$ are non monotonic and 
have the following features:\\
1. A peak  at  $l$ or $d =1$\\
2. A sharp fall at around $l$ or $d=2$ and a subsequent rise.
The fall becomes less steep in time.\\
The feature of a secondary hump is similar to that obtained for
internet and airlines flights networks \cite{gastner}.\\ 
3. Even for the most recent data, the peak at nearest neighbour distances 
is quite dominant. However, with the passage of time, the peak value  
at nearest neighbour distances shrinks while 
the probability at larger distances increases.


\begin{center}
\begin{figure}
\noindent \includegraphics[clip,width= 5cm, angle=270]{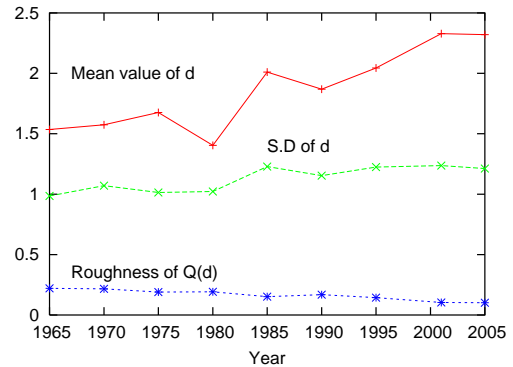}
\caption{ The mean value and standard deviation of
                distances $d$ increase with time while the 
                roughness of the distance distribution $Q(d)$  shows a steady decrease.}
\end{figure}
\end{center}

We have made  a detailed analysis of $Q(d)$, the correlated distance distribution.
In Fig. 4, we present the results.
The mean increases appreciably  in consistency with 
our idea that with the progress of time there will be more collaborations
involving people working at a distance.
The fluctuation  also shows an increase, although its increase
is not that remarkable since the total range of interaction remains fixed in our convention. 
If collaborations were really distance independent, the distributions
$Q(d)$ and $P(l)$ would have looked flat. We have estimated the deviation of $Q(d)$
from a flat distribution by calculating its ``roughness'' $R_Q$  defined as 
$\sqrt{ \langle(Q(d) -\bar Q(d))^2\rangle} $ where $\bar Q(d)$ is the mean value of 
$Q(d)$.
$R_Q$ shows a  decrease with time which is approximately linear. 

The above results imply that even with the communication
revolution, most
collaborations take place among nearest geographical neighbours.
The drop near $d=2$ maybe justified from the fact that in most
cities one has only one university/institute and when one 
collaborates with an outsider, she or he belongs to some other 
city or  country in most cases.
There is some indication that in the not too distant future
collaborations will become almost distance independent as in Fig. 4, 
$R_Q$ seems 
to vanish at around 2040 when extrapolated.
This will mean that the collaboration network takes the nature of
a random network where any two nodes have a finite probability of 
being connected.
It may also happen that $R_Q$ saturates to a finite value in the coming years,
and perhaps it is  too early to predict anything definite.

{\section { Model of collaboration network}}

In this section we  present a  model of the collaboration network
in which spatial and temporal effects are involved.
The aim is to find out the  appropriate scheme by which the links are 
formed in the network such that the observed results are reproduced.
We have taken a two dimensional  space where nodes (authors) can occur 
randomly and each node is assigned
coordinates $x_1,x_2$ where $0\leq x_i\leq 1$. 
Initially we start with a few nodes with a probability $p_0$ to 
have a link with each other.
At each time
step, one new node is introduced. The links are then formed according to the
scheme mentioned below: \\
(a) The new node will 
get  attached to its 
nearest neighbour (Euclidean) with certainty.\\
(b) It  then forms links with probability $p$ to 
the neighbours of its nearest neighbour. \\
(c)  It  also gets attached to  the other  existing nodes 
with probability $q$. \\
In both steps (b) and (c), there can be attachment to more than one
node in general. 
In step (a), distance dependence has been incorporated. 
Step (b) is to ensure that there is a high clustering.
Step (c) has been taken to incorporate the connections 
with neighbours at arbitrary distances.
To keep the model simple, we do not allow new links to form between the 
older nodes. 
Notice that $p_0$ is taken only to
ensure that a connected network is formed and its value should
be kept  small. The distance dependence in this model is 
incorporated by the fact that a new node always gets a link with its 
immediate neighbour. This is motivated by the behaviour of the real data.
(In reality, this may be interpreted as a new research scholar
getting involved in a collaboration with her/his supervisor almost with certainty.)
Since the distance distributions change in time, one must also
incorporate a time dependence in the linking scheme of the model.
$p$ and $q$ are the factors which may
be time dependent. However, in the spirit of the results
obtained, it is reasonable to assume that the time dependence 
of $q$ is more significant and therefore we make  $q$  time dependent
and keep $p$ time independent 
in the minimal model. 
We take $q=q_0t$ where $t$ is the discrete time.
Time dependent probability  to connect to any existing node
has been considered somewhat similarly earlier in a model
of protein interaction network \cite{PIN}.

The time dependence in $q$ ensures two factors: (i) 
as time progresses, a new node gets more links 
and (ii)  since these are   distance-independent, collaborations
with distant neighbours increase with time. 
Both these are close to reality.
In this simple model, we also have only two-author collaborations.
Typically, in most Physics papers, the number of co-authors vary between
2 and 4 \cite{newman1}.   
We have checked from  the real  data   that the distance distribution 
for  papers with two authors is qualitatively very similar to the   
 total distribution (i.e., with any number of authors) 
 and therefore it is sufficient to consider 
 two-author collaborations only in the model network.
  Obviously, $P(l)$ and $Q(d)$ are indistinguishable now.

The choice of $q$ in the form $q=q_ot$ obviously puts a restriction
on the size of the system simulated as $q_0 N$ should be sufficiently less 
than  $1$. Rather than explore the whole parameter space, we have attempted to 
find at least one set of values of $p,q_0$ and $N$ that would be realistic and
also have the observed properties of a collaboration network.
The parameters have been initially chosen such that  the network shows the
proper  spatial and temporal properties of the 
link length distribution, since that was the chief objective of the present
work. Next, we have verified that the other properties are also
well reproduced with this specific choice.
These values are  $q_0 = 10^{-6}$, $p=0.3$ and $N \sim O(10^3)$.
We report the detailed results in the following subsections.\\\\

\subsection{Distance distributions}

In this subsection the simulation results  for the link length distribution
is presented. We also state briefly the reason behind the choice of the
values of the parameters used in the simulations here.

During the evolution of the network, 
the Euclidean distances  between  
each pair of nodes which share a link are  noted.
 In order to verify the behaviour of the distance distribution at 
different times, we have identified a specific number of 
iterations (nodes added) $N_0$ with a test time period, which we call a 
  ``year''.
Here we present the results of the simulation of a network 
with the number of nodes $N= 4N_0$, i.e., for 4 consecutive years. 
For each period of $N_0$ iterations (or one year),
the distance distributions are separately calculated (i.e., not cumulatively) 
to compare
directly with the observed data. 

With both  $p$ and $q$ equal to zero, the distribution would simply 
have a power
law decrease  when nodes get attached to their nearest
neighbours only \cite{mannasen}. 
When $q_0 =0, N = 2000$ and for finite values of $p$ up to approximately  0.5 
the distance distribution shows similarity with the observed data in the
sense that there is a peak at small distances followed by a hump.
This is observed for all the  four sets of $N_0$ iteration steps  
corresponding to the first, second, third and  fourth 
  years (see Fig. 5). However, contrary to the observed data, the 
probability at larger distances does not increase in the later years.
This data, in order to have correspondence with Fig. 2 and 3, should
be a presented in a log-linear plot (as $l \propto \log(D)$).  However, we find that when the 
data is presented in this
way, the probability at larger distances are too small to be visible and
therefore we have shown the log-log plot. This
is another evidence that this model is not the proper one.

\begin{center}
\begin{figure}
\noindent \includegraphics[clip,width= 5cm, angle=270]{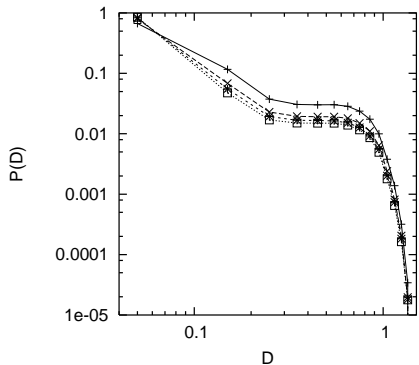}
\caption{
Distance distribution from the simulation data with $q=0$. Here the
data for four  different ``years'' are shown (see text);
earliest year data shown by $+$ and
latest year data by $\Box$.
Distance $D$ is in arbitrary units.  }

\end{figure}

\begin{figure}
\noindent \includegraphics[clip,width= 5cm, angle=270]{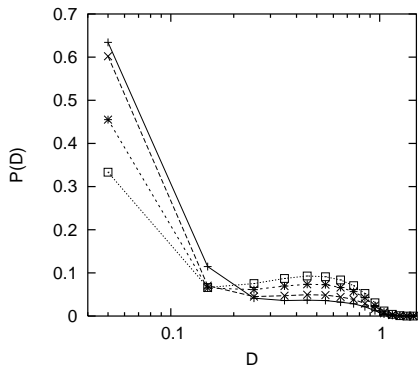}
\caption{
Same as in Fig 5 with $q_0 = 10^{-6}$ presented in a log-linear plot.
Earliest year data are shown by $+$ and
latest year data by $\Box$.  The data  show good agreement with that of
Figs. 2 and 3.
}
\end{figure}
\end{center}

For $p > 0.5$, the distribution becomes too flat to agree with
the observed data for any year. Hence we try with values of  $p$ less than 0.5
and $q_0 \neq 0$. With the same value of $N_0, p=0.3$ and $q_0 = 1.0 \times 10^{-6}$,
we indeed find that the behaviour of the  different  years
reflects the actual behaviour observed in the PRL data (Fig. 6).
Here, one can also present the data in the desired form, i.e., a log-linear
plot with the probabilities at large distances becoming significantly 
larger compared to the case of $q_0=0$.
For    smaller values of $p$, the maximum degree becomes 
much smaller compared to
real world networks. Also, making $q_0$ order of magnitude smaller or 
larger than $10^{-6}$
does not improve the quality of consistency with observed data. 
Having obtained the optimal set of parameter 
values   at  $p=0.3$ and $q_0 = 1.0 \times 10^{-6}$,
we compute the quantities shown in fig. 4, viz.,  
the mean distance, standard deviation and roughness as  functions of time.  
Here slabs of 250 iterations have been taken to correspond to one year to show a larger number of data points and we find that these show excellent 
 agreement with the observed data (Figs. 4 and 7).

\begin{center}
\begin{figure}
\noindent \includegraphics[clip,width= 5cm, angle=270]{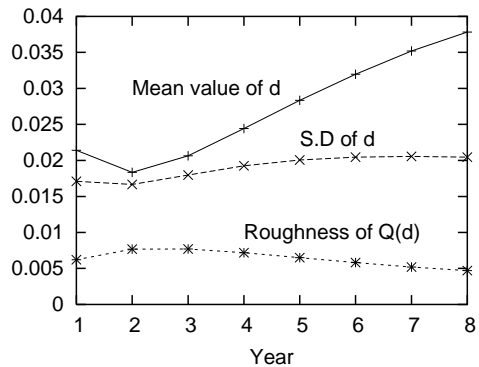}
\caption{
The mean, standard deviation and roughness (see text) shown as functions of 
time (in arbitrary units) calculated  from the simulation. The results may
be compared to that of Fig. 4, obtained from the real database.} 
\end{figure}
\end{center}

\subsection{Small world behaviour}

We have studied the behaviour of   $\langle s\rangle$, the average shortest 
path (i.e., the chemical distance) in the
network as it grows in size.
The behaviour of $\langle s\rangle$ is shown in Fig. 7
as a function of the number of nodes, or equivalently time.
We find that while it is of the order of $\log(N)$ it  shows a non-monotonic
behaviour with $N$. $\langle s\rangle$ increases initially with $N$,
reaches a peak and then decreases with $N$. 
 The decrease with $N$  is consistent with the result in 
\cite{ba_collab} where a
similar result was noted. The decrease maybe attributed to the fact that
the number of edges in the network increases with time in an accelerated
manner such that the average degree increases as the network grows in size.
This is verified by noting the average degree $\langle k\rangle$
as a function of $N$  (Fig. 7). It indeed shows a slow linear 
increase in time agreeing to the behaviour seen in collaboration networks as
reported earlier \cite{ba_collab}.
Note that if we go on increasing the system size,  non-linearity may
occur in the behaviour of $\langle k\rangle$. However, such an increase
is unphysical due to many reasons. This will be discussed in detail in
the following section.
\begin{center}
\begin{figure}
\noindent \includegraphics[clip,width= 5cm, angle=270]{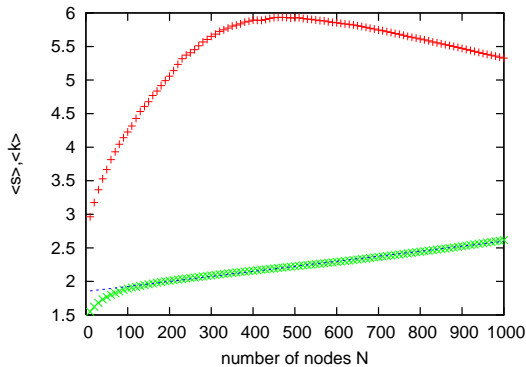}
\caption{
The shortest path $\langle s\rangle$ (upper curve) and  the average degree $\langle k\rangle$ (lower curve)
as functions  of the number of nodes  $N$ (or equivalently time)  are shown.
$\langle k\rangle$ has been fitted to a form $a + bN$, where $a= 1.85$ and 
$b= .75\times10^{-3}$} 

\end{figure}
\end{center}

\subsection{Clustering coefficient}
                                                                                                                             
The clustering coefficient  $\langle C \rangle$
has been calculated by taking the average clustering coefficient of the individual nodes given by
\begin{equation}
C_i= \Sigma_{j_1,j_2} \frac 2{k_i(k_i+1)}{a_{j_1j_2}},
\end{equation}
where $k_i$ is the degree of node $i$ and $a_{j_1j_2}= 1$  if nodes $j_1$ and $j_2$ are connected and zero otherwise.
In social networks the clustering coefficients are usually
quite high, more so for the scientific collaboration network.
Here, however, we do not find such large values as we have restricted
to collaborations between two authors only. However, comparing with the
corresponding random network, we do find that they are much higher.
In Fig. 8, we have shown the variation of $\langle C \rangle$ as a function of time for both the model network and the corresponding 
random network.  It  shows a slow decrease during  later times.
This is again consistent with the results of \cite{ba_collab}.
This may be related to the  observation in \cite{drift}
where it was shown that
while average geographical distance between individual
agents decrease, the group or cluster activity  decreases.
Decrease of clustering coefficient with time implies the tendency of
the network to become random.  However, since in reality, more than two
author collaborations tends to get higher with time this may not actually
be the case after all.
                                                                                                                             
\begin{center}
\begin{figure}
\noindent \includegraphics[clip,width= 5cm, angle=270]{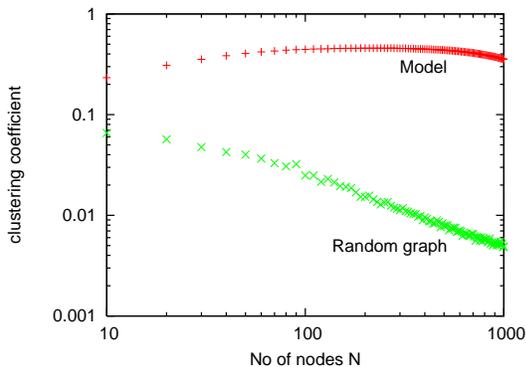}
\caption{
The average clustering coefficient for the present model (upper curve)
and the  corresponding random graph     (lower curve)
as functions  of the number of nodes $N$  (or equivalently time)  are shown.}

\end{figure}
\end{center}
\subsection{Degree Distribution}

 Degree distribution is an important property of networks 
as it determines its behaviour in many respects.

Here we have calculated the degree distribution for the 
simulated model and plotted it in Fig. 10.

 The degree distribution  has a peak, and
is quite similar to that observed in \cite{newman1} for some specific
databases like the cond-mat and hep-th as 
its decay fits to the following form
\begin{equation}
P(k)   \sim k^{-c} \exp(-k/k_0),
\end{equation}
 with $k_0 \sim 5.6$ and $c \sim 1.0$.
The value of $c$ compares very well with that of the observed ones for
several databases (e.g., $c=1.1$ for both cond-mat and hep-th),  
while $k_0$ is fairly close to the values obtained for cond-mat and
hep-th which are 15.7 and 9.4 respectively \cite{newman1}.

\begin{center}
\begin{figure}
\noindent \includegraphics[clip,width= 5cm, angle=270]{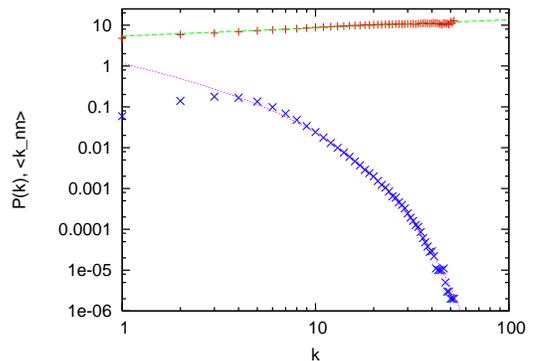}
\caption
{Degree distribution $P(k)$ (lower curve)
 and $\langle k_{nn} \rangle$ (upper curve), the assortativity measure,
  are shown from the simulation of the model.
  $P(k)$ is fit to the form $k^{-1.}\exp(-k/5.6)$ while $\langle k_{nn} \rangle$
  is fitted by $5.4k^{0.2}$ (shown by dashed curves).}
   
\end{figure}     
\end{center}

It maybe mentioned here that
the nature of degree distribution depends largely on the 
particular database under consideration. 
Our results are closer to that of cond-mat or hep-th databases in which the
degree distribution is not a power law. Our model,
which allows only two-author collaborations, is not expected 
to match the data of SPIRES or MEDLINE where collaborations involve a 
large number of researchers and a power-law 
degree distribution has been observed.

\subsection{Assortative mixing}

The assortativity is also another important property of the 
collaboration network. Briefly, it is the degree correlation  of 
nodes on either end of a chosen edge. Here we calculate the assortativity
by calculating the average degree $\langle k_{nn}(k)\rangle$ of the nearest neighbours
of a node with degree $k$. The results show a positive assortativity
as  $\langle k_{nn}(k)\rangle $ clearly increases with $k$ (see Fig. 9).
This is again consistent with real world observations that social networks 
have a positive assortativity \cite{psreview,assort}.

\section{Summary and Discussions}

In this paper, we have reported the results of simulating a scientific 
collaboration network in which both time and space play an important
role in the growth of the model network. The results are compared with the 
observed data of collaboration networks, emphasising on the
link length distribution at different times as this is a feature
not studied in earlier simulations of the collaboration networks. 
The results for the link length distribution agree very  well with the 
observation of the collaborative network of Physical Review Letters presented in section II. 
To test the quality of the model, we have evaluated  other network 
properties for which real data is available in the literature and found 
reasonably good consistency.

This growing network model does not take into account a few features like the `death' of nodes,
change in position of the nodes, more than two author collaborations and collaborations between 
nodes already in the network. These features could easily
be incorporated in the model at the cost of a few new parameters.
We wanted to restrict our model within a few parameters to keep it simple
and yet realistic. Keeping only two node collaborations makes our results
comparable to some specific databases as far as the degree distribution
is concerned (there is an exponential cutoff). All other properties of a collaboration network 
have been successfully reproduced. 

Ignoring death of nodes simply means that generation of the network should
remain limited to finite values of $N$, otherwise the  number of links increases
in a nonlinear manner. Also, $q$ becomes unrealistically high.

New interactions between old nodes would make the number of publications per 
``year'' very high which is not very realistic. A recent study \cite{inertia} shows that
most authors tend to write papers with their old collaborators with more probability, so
that the growth scheme would not be altered much even if one admits such
connections and the results should remain more or less the same.

For small distances, $P(l)$ shows a power law decay contrary to the result
of \cite{katz} where an exponential decay is obtained. Neither the results in \cite{drift} nor our 
results for the PRL data  are  sufficient 
to indicate the exact variation of $P(l)$ with $l$ as distances have been 
coarse grained in both. However, 
a simple argument leads to the conclusion that the small $l$ behaviour of $P(l)$, presented in 
Fig. 2,  may have  a power law decay behaviour.
We notice that there is a sharp decrease of $P(l)$ with $l$ 
for small $l$  which may be assumed to be exponential in nature. 
We have already argued that real
distances  scale roughly as $\exp (\alpha l^\beta)$ where $\beta$ is  of
the order of unity.
In that case, the initial exponential decay of $P(l)$ with $l$ 
corresponds to a power law decrease with  true distances $D$. 
In \cite{katz}, the data base was differently generated which may be the 
cause of the discrepancy in the behaviour of $P(D)$ for small $D$. 
Our simulations also show a sharp decrease of $P(D)$ with $\log(D)$.

It may be mentioned here that for the real world data, we have coarse
grained the distance (according to cities, countries, continents etc.) while no such scheme was taken up for the model.
Even then, the link length distribution from the simulation shows 
reasonably good agreement with the data. The reason 
is  that a new node in the simulation invariably links up with its nearest node, and this nearest node
is expected to lie sufficiently `close'  to it  (nodes exist randomly all over the 
space) 
so the distribution $P(D)$ at small distances gets 
enriched. 
The existence of the `dip' in $P(D)$ in the simulation result 
can be explained in the following way: 
we have connections  with neighbours at arbitrary  distances through
steps (b) and (c) (in fact step (c) contributes more towards this). 
In a Euclidean space, the number of points lying  within a shell of 
thickness $dr$ at a distance $r$ is $2\pi r dr$ in two dimensions.
	Naturally, the number of such points increases with $r$, and  
   therefore 
 an  increase in $P(D)$ at greater  $D$ is possible 
 resulting in a dip in between.

In generating the network here, we could have assumed a form of $P(D)$ as has been done earlier \cite{ps_rev}
rather than using the  scheme described in the beginning of  section III.
But we have not attempted to do this for two reasons \\
1. The exact form of $P(D)$ is unknown, neither does it seem to be a simple one.\\
2. The present scheme, being successful, helps to develop newer insight in the
evolution of the collaboration  network. 

The question may now arise that whether  ignoring
the distance dependence while constructing a model of collaboration
network is justified, which has been done in earlier works.
This leads to an intriguing realisation. In the present  model, the distance dependence 
matters for  nearest neighbours only.
Indeed,  even in   the present   model one can ignore the distance
factor when  a different perspective is taken. Instead of
assuming that a new node is born randomly at any position and that it gets 
connected to its nearest neighbour,
one may suppose that a random node (i.e., the  existing nearest neighbour
or the parent) 
has duplicated and the
duplicated node (daughter) is always connected to the 
parent. 
The daughter  
node is  also connected with 
probability $p$
to its parent's neighbours and with $q$ to others. 
This is then simply a non-Euclidean network!
Obviously this equivalent non-Euclidean model does not carry
any information of the distance dependence but would give us the same
values of shortest paths, clustering coefficients, degree distribution
and assortativity.

Acknowledgments: We thank
the fellow members of the network group of Calcutta 
for helpful discussions.
We   acknowledge support from CSIR grants  03(1029)/05/EMR-II (PS),
9/28(609)/2003-EMR-I (KBH) and
9/28(608)/2003-EMR-I (PD).
AKC is grateful to support from UGC.
Financial support from DST FIST for computational work is also
acknowledged.


\end{document}